\documentstyle[epsfig,twocolumn,aps]{revtex} 

\begin{document}
\pagestyle{myheadings}
\markright{\today}
\def\lapprox{{\raise0.5ex\hbox{$<$}\hskip-0.80em\lower0.5ex\hbox{$\sim$}
}}
\def\gapprox{{\raise0.5ex\hbox{$>$}\hskip-0.80em\lower0.5ex\hbox{$\sim$}
}}
\draft

\title{
The Reaction $^7$Li$(\pi^+,\pi^-)^7$B and its Implications for $^7$B
\thanks{supported by the BMBF (06 TU 886), DFG (Mu 705/3,
Graduiertenkolleg), NFR and INTAS RFBR (95-605)}
}
\author{
J. P\"atzold, R. Bilger, H. Clement, K. F\"ohl\thanks{Present address: Department
of Physics and Astronomy, University of Edinburgh}, J. Gr\"ater, \\ 
R. Meier, D. Schapler\thanks{Present address: SAP, Walldorf, Germany} and G.J. Wagner}
\address{
Physikalisches Institut der Universit\"at T\"ubingen, \\  Auf der
Morgenstelle 14, D-72076 T\"ubingen
}
\author{A. Denig and W. Kluge}
\address{
Institut f\"ur Experimentelle Kernphysik, Universit\"at Karlsruhe
}
\author{M. Schepkin}
\address{
Institute for Theoretical and Experimental Physics, Moscow
}

\maketitle

\begin{abstract}
The reaction $^7$Li$(\pi^+,\pi^-)^7$B has been measured at incident
pion energies of 30--90 MeV. $^7$Li constitutes the lightest target
nucleus, where the pionic charge exchange may proceed as a binary
reaction to a discrete final state. 
Like in the $\Delta$-resonance region the observed cross sections are much
smaller than expected from the systematics found for heavier nuclei. 
In analogy to the neutron
halo case of $^{11}$Li this cross section suppression is interpreted
as evidence for a proton halo in the particle-unstable nucleus $^7$B.
\end{abstract}

\vspace{\bigskipamount}

\narrowtext

In recent years the pionic double charge exchange (DCX) reaction  
has received much
attention  at energies below the $\Delta$ resonance,
mainly  for two reasons. On the one hand the cross
sections there have been found to be sensitive to nucleon-nucleon
(NN) correlations of short range, a feature which has been looked for
since long in this genuine 2N reaction \cite{joh93}. On the other
hand the forward-angle cross sections exhibit an unexpected though
systematic resonance-like energy dependence with peak cross sections
between $T_\pi = 45$ and 70 MeV, which in general are substantially
larger than those in the $\Delta$-resonance region and above.
Combination of both these features has led to the so-called $d'$
hypothesis \cite{bil93,foe97}, which postulates the existence of a
NN-decoupled $\pi$NN resonance with $I(J^P) =$ even $(0^-), m \approx
2.06$ GeV and vacuum width $\Gamma_{\pi{\rm NN}} \approx 0.5$ MeV to
explain the observed effects in DCX. In the nuclear medium the width
of such a resonance is broadened very much by ``Fermi smearing'' due
to the motion of the nucleon pair active in the DCX process as well
as by collision damping due to $d'N \to 3N$, so that the resulting
width in the nuclear medium lies in the order of 20 to 30 MeV. The
question whether this picture is correct or whether a subtle, as of
yet not understood medium effect is the origin of this resonance-like
structure, is not easy to settle within the DCX, since the free
process on a dinucleon is not observable. However, one could expect
to minimize the influence of such contingent medium effects by
studying the DCX process on the lightest nuclei possible. $^7$Li is
the lightest nucleus where this reaction may still proceed to a
discrete final nuclear state, though the $^7$B ground state is
already 3.65 MeV above the proton emission threshold. As a result it
has a natural width of 1.4(2) MeV \cite{ajz88} corresponding to a
lifetime of $5 \times 10^{-22}$ s. This is still in the order of
magnitude of the classical orbiting time for the 3 valence protons
outside the alpha core in $^7$B. Hence it appears justified to ask
about the radius of these proton orbits, in particular whether they
possibly form a proton halo. A comparable case  has recently been discussed
regarding proton-unstable states in $^{17}$F \cite{mor97}.
With regard to DCX a similar situation on the neutron side has
recently been met in the $^{11}$B$(\pi^-,\pi^+)^{11}$Li reaction,
where it has been demonstrated \cite{gib91} that this reaction is
extremely sensitive to the neutron halo in $^{11}$Li and provides a
reliable determination of its radius. 

\begin{figure}
\centerline{
\epsfxsize0.5\textwidth
\addtolength\epsfxsize{-1cm}
\epsfbox{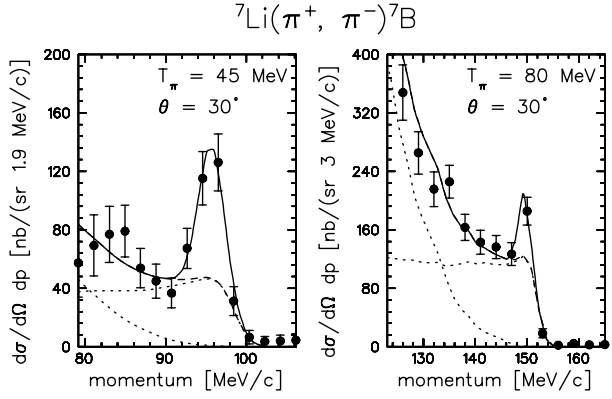}}
\bigskip
\caption[]{Sample DCX spectra taken at $T_\pi = 45$ and 80 MeV, respectively,
at a scattering angle of $\Theta_{lab} = 30^\circ$ .
The dotted lines represent
$^6$Be + p and $^4$He + 3p breakup channels. The transition to the
$^7$B ground state is fitted by a Gaussian, the width of which has
been determined from the experimental resolution measured in
corresponding elastic scattering runs.
\label{fig1}}
\end{figure}

The measurements have been carried out with the LEPS magnetic
spectrometer \cite{bar90} at the $\pi$E3 channel at PSI. 
Sheets of metallic Lithium with an areal density of 265
mg/cm$^2$ and a $^7$Li isotopic purity of about 95\% served as target. Absolute 
cross
sections have been obtained by use of the lepton normalization method
\cite{bar90} which is based on the measurement of elastic $\mu$
scattering. The simultaneously measured elastic $\pi$ scattering
serves as cross check for the validity of this method. The cross
sections obtained this way for elastic $\pi$ scattering from $^7$Li
are in very good agreement with optical model calculations using the
J4 potential \cite{mei89}, which is known to provide reliable predictions for
low pion energies. Sample spectra obtained for the DCX reaction at
$\Theta_{lab} = 30^\circ$ are shown in Fig. \ref{fig1} for low and high
incident energies. In both spectra the peak corresponding to the
transition to the $^7$B ground state (GST) is clearly visible,
however, the continuum due to the breakup channels $^6$Be + p and
$^4$He + 3p with  Q-values of $+2.28$ and $+3.65$ MeV,  respectively,
relative to the $^7$B ground state increases strongly with increasing
incident energy. Hence especially for the higher energies a proper
treatment of the continuum is important for a reliable extraction of
the peak content. In ref. \cite{set92} it has been shown that the
continuum due to breakup of light nuclei is not well described by
pure phase space, in particular at its high-energy end, where the
breakup fragments have very low relative energies and hence are
likely to undergo substantial final state interactions (FSI).
Therefore we have taken into account FSI \cite{sch93} in the calculation of the
shape of the breakup continuum. It is significant particularly for
the two-body breakup $^6$Be + p which starts with a steep rise right
at the position of the GST peak. Fig. \ref{fig1} shows the decomposition of
the experimental data into continuum part (dashed lines) and GST
peak. The description of the continuum part is accomplished by
adjusting the absolute magnitude of the calculated distributions for
the two breakup channels (dotted) to the data. The GST peak is
described by a Gaussian the width of which has been calculated from
the experimental resolution obtained in the corresponding elastic
scattering runs.

\begin{figure}
\centerline{
\epsfxsize0.5\textwidth
\addtolength\epsfxsize{-1cm}
\epsfbox{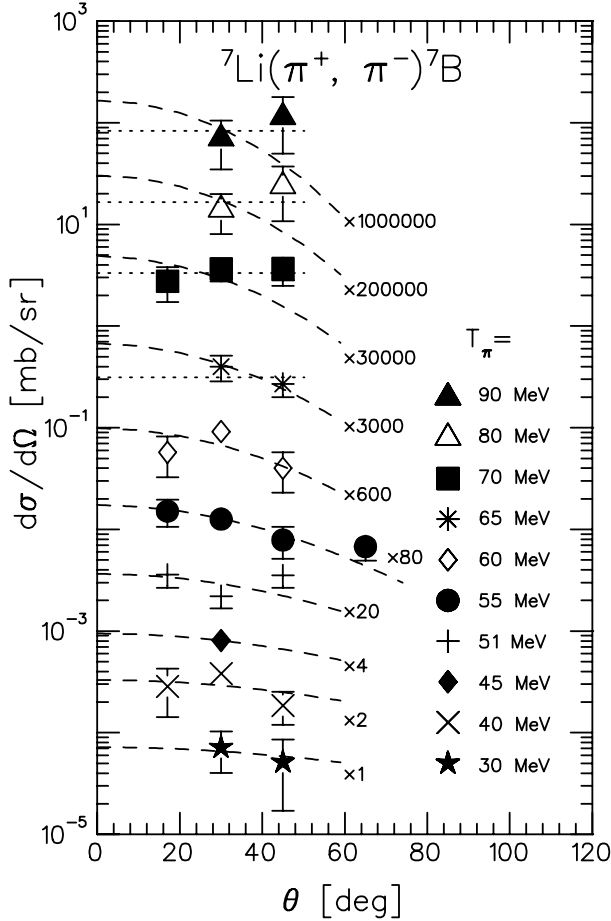}}
\bigskip
\caption[]{Angular distributions of the GST in the energy range $T_\pi =
30$ - 90 MeV. The dashed curves represent $d'$ calculations fitted in
height to the data. The horizontal lines characterize an isotropic
angular dependence fitted to the data for $T_\pi \geq 65$ MeV.
\label{fig2}}
\end{figure}

The experimental angular distributions  for the GST 
are shown in Fig. \ref{fig2} for the different incident energies. The error
bars include the uncertainties both from statistics and from the
decomposition of the spectra into GST peak and breakup continuum. The
observed angular dependence is rather weak as expected for monopole
transitions on light nuclei. This is also borne out by the
calculations (dashed curves, adjusted in height to the data)
performed within the $d'$ model, which will be discussed below in
more detail. Since in this model the angular dependence is governed
by the c.m. motion of the active nucleon pair in initial and final
nuclear states, i.e. by the two-nucleon form factor, the predicted
angular dependence equals essentially the one expected  from a standard
DCX treatment. The data are compatible with the predicted angular
dependence up to $T_\pi \approx 60$ MeV. For the higher energies 
$T_\pi~\gapprox~70$ MeV, however, where the calculations predict an
increasing fall-off towards larger 
angles, the data exhibit a nearly
opposite behavior with cross sections tending to increase with
increasing angle. This change in the experimental angular behavior is
clearly borne out by comparing the 55 MeV and 70 MeV data sets. Since
at high incident energies the GST peaks sit upon a large continuum of
breakup channels, the question might arise whether some incorrect
treatment of the latter might have affected the GST data.
However, we tried various descriptions of the background and
none led to significant changes in the GST angular
dependence. If there is no experimental problem with the data at
these higher energies, then this change in the angle dependence has
to be associated with some physical origin. As will be discussed
below, beyond 60 MeV the $d'$ amplitude gets already small compared
to the conventional amplitude, which we identify with the tail of the
$\Delta\Delta$ process \cite{joh93}, so that the angular
distributions no longer need to be governed by the $d'$ mechanism. We
only note in passing that in the DCX on still lighter nuclei, the He
isotopes, the angular distributions for incident energies between 70
and 120 MeV are observed to be practically isotropic, too
\cite{yul97,gra98}.

\begin{figure}
\centerline{
\epsfxsize0.5\textwidth
\addtolength\epsfxsize{-1cm}
\epsfbox{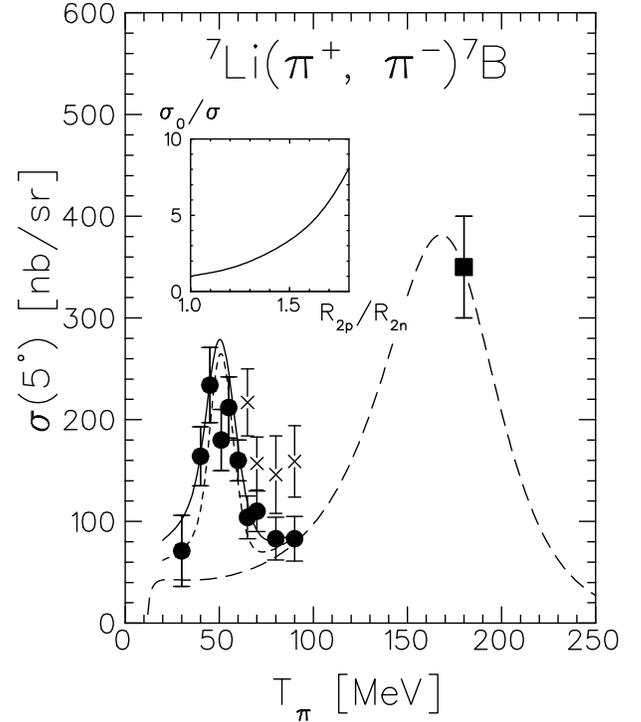}}
\bigskip
\caption[]{Energy dependence of the forward angle cross section $\sigma
(\Theta = 5^\circ)$. The data point at $T_\pi = 180$ MeV is from ref.
\cite{set90}. At low energies our extrapolated values as obtained
from Fig. \ref{fig2} are shown. For $T_\pi \geq 65$ MeV we give two
values: for the solid dots an isotropic angular distribution is
assumed as suggested by the data, for the crosses the validity of the
$d'$ angular distributions is assumed though they miss the trend in
the data. The dashed lines represent the $\Delta\Delta$ process in a
phenomenological parametrization. The dotted and solid curves give
the result, when the $d'$ amplitude is added coherently using
$\Gamma_{spread} = 10$ and 15 MeV, respectively. The inset shows the
change in the calculated cross section due to a change in the orbit
radius of the active NN pair (see text).
\label{fig3}}
\end{figure}

Fig. \ref{fig3} displays the energy dependence of the forward angle cross
section. For the $\Delta$ resonance region there exists one
measurement at $T_\pi = 180$ MeV and $\Theta_{lab} = 5^\circ$
\cite{set90}. In order to compare with its result we have
extrapolated our data to $\Theta_{lab} = 5^\circ$ by the curves shown
in Fig. \ref{fig2}. Since the angular distributions are very flat these
extrapolations are quite moderate except at the highest energies.
Since there the observed angular dependence is at variance with the
$d'$ predictions we give two values in Fig. \ref{fig3} for $T_\pi \geq 65$
MeV, the one obtained with the $d'$ angular dependence as for the
lower energies (crosses), and the one obtained simply by assuming an isotropic
angular dependence as suggested by the data (dots). 

\begin{figure}
\centerline{
\epsfxsize0.5\textwidth
\addtolength\epsfxsize{-1cm}
\epsfbox{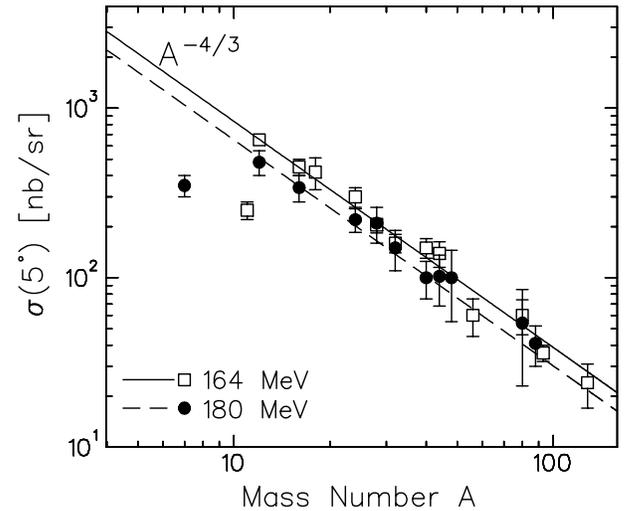}}
\bigskip
\caption[]{Systematics of the GSTs in the $\Delta$-resonance region.
Open (solid) symbols represent the data at $T_\pi = 164(180)$ MeV
(from refs. \cite{set90,gil87,kob92}). The solid and dashed lines
give the $A^{-4/3}$ dependence fitted to these data.
\label{fig4}}
\end{figure}

The most conspicuous and common feature of both the LAMPF datum and our
measurements is the surprisingly small cross sections, which are much
smaller than expected from systematics. For the discussion of this
issue we first turn to the $\Delta$ resonance region, where a great
deal of systematic DCX studies have been undertaken. There both
analog and nonanalog transitions have been shown \cite{joh93,gil87}
to exhibit very simple systematic dependences on the target mass A,
which may be understood in simple diffractive models of the DCX process
in the $\Delta$  region. Fig. \ref{fig4} shows as an example the forward angle data
\cite{set90,gil87,kob92} for the nonanalog GSTs at $T_\pi = 164$ and
180 MeV, respectively. At both energies the data are in excellent
agreement with the expected $A^{-4/3}$ dependence with the exception
of two cases, the $^{11}$B$(\pi^-,\pi^+)^{11}$Li cross section \cite{kob92}
at 164 MeV and the $^7$Li$(\pi^+,\pi^-)^7$B cross section \cite{set90} at 180
MeV. Both are roughly a factor of three below their value expected
from the $A^{-4/3}$ systematics. For the $^{11}$B case this huge
discrepancy has recently been successfully explained by the neutron
halo of $^{11}$Li. According to Gibbs and Hayes \cite{gib91} GSTs are
very sensitive to a change of the active nucleons' orbital radius in
the transition from initial to final states. In the limiting case of
extreme halo radii the cross section is expected to scale as the
inverse sixth power of the halo radius. For realistic cases this
dependence is much more moderate. In the case of
$^{11}$B$(\pi^-,\pi^+)^{11}$Li the  ratio of the measured cross section
$\sigma$ over the value $\sigma_0$ expected from systematics is 
$1/3.3 (4)$. This ratio has been related in ref. \cite{gib91} to a change of the
radius of the active proton pair $R_{2p}$ in $^{11}$B to that of the active
neutron pair $R_{2n}$ in $^{11}$Li by $R_{2n}/R_{2p} = 1.9(3)$. This
represents rather a quadratic dependence. The latter uncertainty
includes also systematic uncertainties in the calculations (wave
functions, distortion etc.). The resulting halo radius of $R_{2n} = 5$ fm for
$^{11}$Li determined this way agrees very well with that obtained by
other means \cite{kob92}. 

It is therefore very tempting to try to
explain the surprisingly low cross sections for
$^7$Li$(\pi^+,\pi^-)^7$B in full analogy to the above example. In
this case  we would rather deal with a proton halo in $^7$B, which
appears to be not unplausible in view of the particle-unbound
character of this nucleus, as discussed already in the introduction.
From Fig. \ref{fig4} we see that in this case the suppression is 
 $\sigma_0/\sigma = 2.9(4)$,
which translates into $R_{2p}/R_{2n} = 1.6(3)$, if we transfer the
findings for $^{11}$Li (Fig. 1 of ref. \cite{gib91}) straight to our case.
However, we have to be a bit more careful, since in contrast to $^{11}$B the
target nucleus $^7$Li has already an unusually large charge radius due to its
loosely bound $\alpha + t$ cluster structure. With 2.4 fm its root-mean-square radius is
roughly 15\% larger than expected from the systematics of nuclear radii. Since
the $A^{-4/3}$ dependence of forward angle cross sections derived 
in diffractive models results from
geometrical considerations and represents essentially a $R^{-4}$ dependence,
the increased $^7$Li radius causes $\sigma_0$ already to be lower by a factor of
 about
1.7. This leads then to a reduced suppression of $\sigma_0/\sigma = 1.7(2)$, which
translates into a value of $R_{2p}/R_{2n} = 1.3(2)$ by Fig. 1 of ref. 
\cite{gib91}.

As pointed out in ref. \cite{gib91} the reduction factor in cross
section due to the change of $R_{2N}$ is largely independent of the
reaction mechanism and results primarily from the overlap of the
initial and final NN-wave functions. Hence also the $d'$ model, which contains
this overlap, too, and which
we will use for the description of our low-energy data, should be
adequate for estimating $R_{2p}/R_{2n}$, even from the datum at $T_\pi = 180$
MeV. The function $\sigma_0/\sigma = f(R_{2p}/R_{2n})$ calculated
this way is shown as inset in Fig. \ref{fig3}, it yields for the ratio of
$\sigma_0/\sigma = 1.7(2)$ a value of $R_{2p}/R_{2n} = 1.2 - 1.3$ which
agrees very well with the value obtained above.

With this in mind we now turn to the discussion of the low-energy
data. The first striking feature there is that the peak cross section is
much below the cross section at the $\Delta$ resonance, a feature
observed so far in no other case. Even for the neighboring cases
$^{12}$C and $^{16}$O \cite{foe97} the low-energy peak cross sections
are still well above those in the $\Delta$-resonance region. Since
the low-energy cross sections are known to depend strongly  on NN
correlations, it is tempting to seek the solution there. Indeed,
the unusually large charge radius of $^7$Li
leads to a considerable reduction of the probability to find the
valence neutrons active in the DCX process at small relative
distances. In the $d'$ calculations this alone gives  a reduction by
roughly a factor of two in cross section compared to the case, where
the usual $A^{1/3}$ dependence for the radius of $^7$Li is assumed.
Another strong reduction in the $d'$ cross section results, if in addition  for
$^7$B a proton halo is assumed as discussed in the preceding section.
The cross sections calculated this way are displayed in Fig. \ref{fig3}. The
dotted curve shows the result, if we use $R_{2p}/R_{2n} = 1.3$, as
derived from the 180 MeV datum, and a value of $\Gamma_{spread} = 10$
MeV for the collision damping. These calculations account already
quite well for the data, only the calculated resonance structure
appears to be somewhat too narrow. This can be improved if we use
$\Gamma_{spread} = 15$ MeV instead, but then the calculated peak
cross section gets somewhat too low. The calculations may be brought
back to the data if we readjust the radius ratio to $R_{2p}/R_{2n} =
1.2$ (solid curve in Fig. \ref{fig3}). So our conclusions about a proton
halo in $^7$B are somewhat dependent on
$\Gamma_{spread}$. The data clearly prefer the larger value for
$\Gamma_{spread}$ and hence the more moderate value for the change
in radii. Yet, the value $R_{2p}/R_{2n} = 1.2$  is still in good agreement 
with the one derived above from
the cross section in the $\Delta$-resonance region.

In conclusion, the measured cross sections for the DCX on $^7$Li
leading to the ground state in $^7$B are much smaller than expected
from systematics both in the $\Delta$ resonance region and below.
This suppression can  partly be attributed to the exceptionally large radius
of $^7$Li and partly to an even larger  proton halo
in $^7$B with a radius of about 3  fm, i.e. as big as the radius of nuclei in the
Ca region. The low-energy data show again some peak structure in the
energy dependence, though much less pronounced than observed for
other nuclei. Within the $d'$ model this suppression is understood as
being due to the low NN separation densities at short relative
distances because of the large radii in $^7$Li and in particular in
$^7$B. The observed angular dependence is well understood at
energies, where the $d'$ amplitude is the dominating process. At
higher energies, where the $d'$ amplitude is small compared to other
processes, the experimental angular distributions get surprisingly
flat, a phenomenon not yet understood.

\begin{references}
\bibitem{joh93}  for a survey see, e.g. M.B. Johnson and C.L. Morris,
Ann. Rev. Nucl. Part. Sci. {\bf 43} (1993) 165; H. Clement, Prog.
Part. Nucl. Phys. {\bf 29} (1992) 175; and references therein
\bibitem{bil93} R. Bilger, H.A. Clement and M.G. Schepkin, Phys. Rev. Lett.
{\bf 71} (1993) 42 and {\bf 72} (1994) 2972 
\bibitem{foe97} K. F\"ohl et al., Phys. Rev. Lett. {\bf 79} (1997) 3849
\bibitem{ajz88} see F. Ajzenberg-Selove, Nucl. Phys. {\bf A490}
(1988) 1
\bibitem{mor97} R. Morlock et al., Phys. Rev. Lett. {\bf 79} (1997) 3837
\bibitem{gib91} W.R. Gibbs and A.C. Hayes, Phys. Rev. Lett. {\bf 67}
(1991) 1395
\bibitem{bar90} B.M. Barnett et al., Nucl. Instr. Meth. {\bf A297}
(1990) 444; H. Matth\"ay et al., Proc. Int. Symp. on Dynamics of
Collective Phenomena in Nuclear and Subnuclear Long Range
Interactions in Nuclei, Bad Honnef 1987 (ed. P. David) World
Scientific 1988, 542
\bibitem{mei89} O. Meirav et al., Phys. Rev. {\bf C40} (1989) 843
\bibitem{set92} K.K. Seth, Int. Workshop on Pions in Nuclei,
Penyscola 1991 (eds. E. Oset, M.J. Vicente-Vacas, C. Garcia Recio)
World Scientific 1992, 205
\bibitem{sch93} see, e.g. M.L. Goldberger and K.M. Watson, Collision Theory
  (John Wiley \& Sons, New York, 1964);
M. Schepkin, O. Zaboronsky, H. Clement,
  Z. Phys. {\bf A345} (1993) 407
\bibitem{yul97} M. Yuly et al., Phys. Rev. {\bf C55} (1997) 1848
\bibitem{gra98} J. Gr\"ater et al., Phys. Lett. {\bf B420} (1998) 37 and
  Phys. Rev. {\bf C}, accepted for publication
\bibitem{set90} K.K. Seth, Second LAMPF Int. Workshop on Pion-Nucleus
Double Charge Exchange, Los Alamos 1989 (eds. W.R. Gibbs and M.J.
Leitch) World Scientific 1990, 473
\bibitem{gil87} R. Gilman et al., Phys. Rev. {\bf C35} (1987) 1334
and references therein
\bibitem{kob92} T. Kobayashi, Nucl. Phys. {\bf A538} (1992) 343c and
{\bf A553} (1993) 465c
\end{references}
\end{document}